\documentclass[aps,prl,twocolumn,showpacs,superscriptaddress,nofootinbib,preprintnumbers]{revtex4-2}

\usepackage{amsmath}
\usepackage{graphicx}
\usepackage{dcolumn}
\usepackage{bm}
\usepackage{hyperref}
\usepackage{color}
\usepackage{xcolor}
\usepackage{soul}
\usepackage{slashed}
\usepackage{tikz}
\usepackage{multirow}
\usepackage{mathrsfs}
\usepackage{cleveref}
\usepackage{comment}
\usepackage{ragged2e}
\usepackage{ulem}

\hypersetup{
    colorlinks,
    linkcolor={red!60!black},
    citecolor={blue!60!black},
    urlcolor={red!60!black}
}

\newcommand{\be}{\begin{eqnarray}}
\newcommand{\ee}{\end{eqnarray}}
\newcommand{\beq}{\begin{equation}}
\newcommand{\eeq}{\end{equation}}
\newcommand{\Ap}{ {A^\prime} }

\newcommand{\brac}[2]{ \left( \frac{#1}{#2} \right) }

\definecolor{wildstrawberry}{rgb}{1.0, 0.26, 0.64}

\definecolor{puke}{rgb}{0.7, 0.7, 0.4}

\begin{document}

\preprint{FERMILAB-PUB-25-0234-T}

\title{Testing Thermal-Relic Dark Matter with a Dark Photon Mediator
}

\author{Gordan Krnjaic}
\affiliation{Theory Division, Fermilab, Batavia, IL, USA}
\affiliation{Department of Astronomy and Astrophysics,  Kavli Institute for Cosmological Physics \\  
University of Chicago, Chicago, IL, USA}

\date{\today}
\begin{abstract}
In light of recent DAMIC-M results, we present the status of thermal-relic dark matter $\chi$ coupled to a kinetically-mixed dark photon $A^\prime$.
In the predictive ``direct annihilation" regime, $m_{\Ap} > m_\chi$, the relic abundance depends on the kinetic mixing parameter, and there is a minimum value compatible with thermal freeze out.
Using only electron and nuclear recoil direct detection results, we find that for complex scalar dark matter, the direct-annihilation regime is now excluded for nearly all values of $m_\chi$; the only exception is the resonant annihilation regime where $m_\Ap \approx 2 m_\chi$.
Direct annihilation relic targets for other representative models, including Majorana and Pseudo-Dirac candidates, remain viable across a wide range of model parameters, but will be tested with a combination of dedicated accelerator searches in the near future. 
 In the opposite ``secluded annihilation" regime, where $m_\chi > m_\Ap$, this scenario is excluded by cosmic microwave background measurements for all $m_\chi \lesssim 30$ GeV. Similar conclusions in both the direct and secluded regimes hold for all anomaly-free vector mediators that couple to the first generation of electrically-charged Standard Model particles. 
\end{abstract}

\maketitle


\section{Introduction}
Thermal freeze out is the only  
dark matter (DM) production mechanism that is insensitive to the initial conditions of our universe \cite{Cirelli:2024ssz}. In this framework, 
DM maintains chemical equilibrium with Standard Model (SM) particles
until annihilation reactions become inefficient relative to Hubble expansion and the comoving DM density is conserved. If DM annihilates directly to SM final states, there is a minimum DM-SM coupling compatible with thermal production, and a corresponding experimental target for discovering or falsifying this compelling scenario.

Although freeze out has historically been associated with DM masses near the electroweak scale \cite{Bertone:2016nfn}, under standard cosmological assumptions, the mechanism is compatible with masses anywhere in the  MeV-100 TeV range.\footnote{If DM-neutrino equilibrium is only established after neutrino decoupling, the mass range can extend down to the keV-scale \cite{Berlin:2018ztp,Berlin:2019pbq}. If additional entropy transfers occur after freeze-out, the mass range can extend beyond  $\sim$100 TeV \cite{Berlin:2016gtr,Berlin:2016vnh}} For DM below the GeV scale, freeze out via direct annihilation to SM particles requires comparably-light new mediators, otherwise the suppression from weak scale masses in the annihilation rate results in DM overproduction \cite{Lee:1977ua}. Since such light mediators must be neutral under the SM gauge group and must also couple to visible matter through renormaliazble  ``portal" operators -- the vector portal \cite{Fabbrichesi:2020wbt}, the Higgs portal \cite{Krnjaic:2015mbs}, the neutrino portal \cite{Batell:2017cmf,Bell:2024uah} -- there is a relatively short list of viable options for these particles.

\begin{figure*}[t!]
\centering
\includegraphics[width=0.9\linewidth]{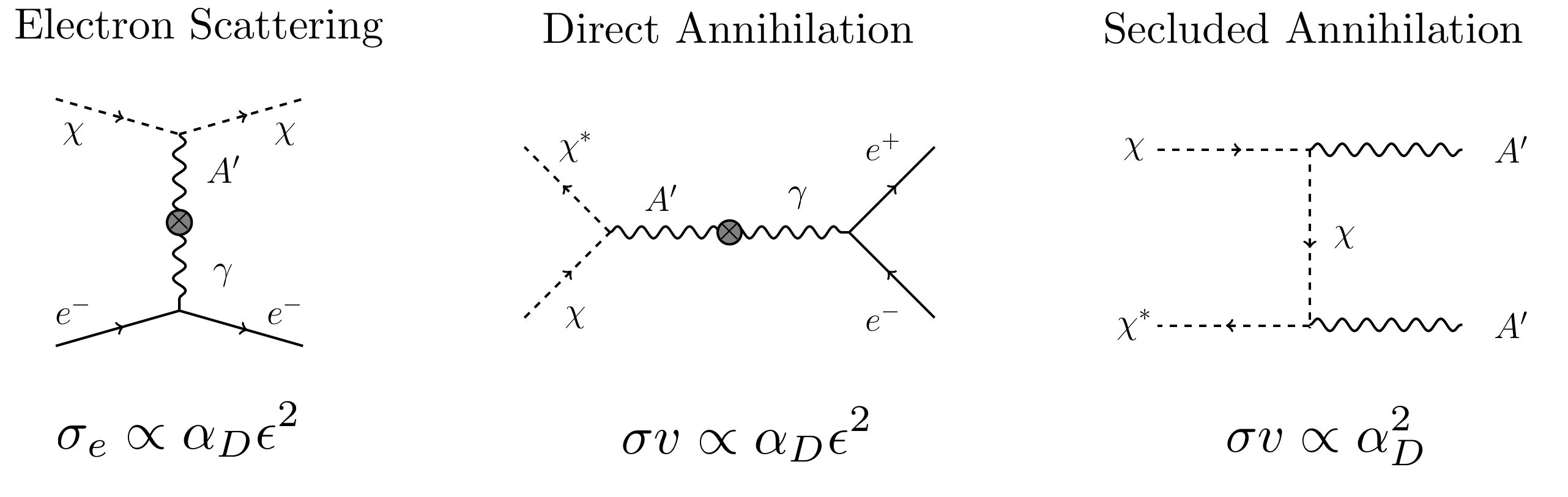}
\vspace{-0.cm} f
\caption{Representative Feynman diagrams that depict the processes studied in this work. {\bf Left}: dark matter scattering off electrons. {\bf Middle}: DM direct annihilation  to SM particles through the kinetic mixing interaction. This process is always kinematically allowed for DM masses compatible with a thermal freeze out origin. For $m_\chi < m_\Ap$, this is the only allowed annihilation channel and it sets the relic density. 
{\bf Right:} In the absence of other interactions, for $m_\chi > m_\Ap$, the relic density is governed by  the ``secluded" annihilation channel which is independent of the $\Ap$-SM coupling $\epsilon$, so there is no experimental target. However, the freeze out mechanism is excluded in the this regime for $m_\chi < 10$ GeV due to CMB limits on energy injection 
}
\label{fig:cartoon}
\end{figure*}

In this {\it Letter}, we present the status of thermal-relic dark matter whose interactions with the SM are mediated through a massive dark photon $\Ap,$ with the interaction Lagrangian
\be
{\mathscr L}_{\rm int} = -  A^\prime_\mu(  g_D J_D^\mu +   \epsilon e J_{\rm EM}^\mu  ),
\ee
where $g_D = \sqrt{4\pi \alpha_D}$ is the dark gauge coupling, and $\epsilon$ is the kinetic mixing parameter that governs interactions with the electromagnetic current, $J^\mu_{\rm EM}$ \cite{Fabbrichesi:2020wbt}. 
In predictive ``direct annihilation" models (Fig. \ref{fig:cartoon}),  the relic density is in one-to-one correspondence direct-detection cross section and we consider the benchmark models
\be
\label{eq:models}
J_D^\mu = 
\begin{cases}
~~i\chi^* \partial^\mu \chi + c.c. & \text{Scalar}  
\\
~~~~~\frac{1}{2}\overline \chi \gamma^\mu\gamma^5 \chi  & \text{Majorana}
\\
 ~~~~~~ i \overline \chi_1 \gamma^\mu \chi_2 & \text{Pseudo-Dirac} 
 \\
 ~~~~~~~\overline \chi \gamma^\mu \chi  & \text{Dirac (Asymmetric)}
\end{cases}
\ee
which feature direct-annihilation targets for thermal production. In the Pseudo-Dirac case, DM consists 
of two nearly mass-degenerate states $\chi_{1,2}$ with an inelastic coupling to $\Ap$. In the degenerate limit, these states form a Dirac fermion, but in order for this model to be viable, there must be a particle-antiparticle asymmetry (see below). Nonetheless, avoiding overproduction requires a minimum annihilation cross section, so this model still has a predictive discovery target.

Taking into account recent DAMIC-M results \cite{DAMIC-M:2025luv},  we find that the scalar scenario is now fully excluded by a combination of electron- and nuclear-recoil direct-detection bounds. The only exception is the fine-tuned resonance region $m_\Ap \approx 2 m_\chi$, and similar considerations apply to the asymmetric Dirac model, which also predicts appreciable direct-detection rates in the thermal production regime. For the scalar model, this finding updates the conclusion of Ref. \cite{Balan:2024cmq} in light of decisive new data. 
By contrast, the Majorana and Pseudo-Dirac models predict sharply suppressed scattering cross sections, so these are currently beyond the reach of existing direct-detection
experiments. However, future $B$-factory and fixed-target accelerator experiments can comprehensively cover these targets with dedicated searches in the near future.

\section{Thermal Relic Targets}
\noindent {\bf Freeze Out Formalism.} In the early universe, $\chi$ is in equilibrium with the SM  and the total DM number density $n_\chi$  evolves as 
\be
\dot n_\chi + 3 H n_\chi  = -k \langle \sigma v \rangle [
 n_\chi^2 - (n^{\rm eq}_\chi  )^2  ],~~~
\ee
where $H$ is the Hubble rate, $n_\chi^{\rm eq}$ is the $\chi$ number density in chemical equilibrium with the SM, and $k = 1$ or $1/2$ for identical or non-identical annihilating particles, respectively.
The thermally-averaged cross section can be written \cite{Gondolo:1990dk}
\be
\langle \sigma v \rangle   =   \frac{1}{8m^4_\chi T K^2_{2} \! \brac{ m_\chi}{T}} \!  \int^\infty_{s_0}
\! \! \! ds \,
\sigma(s) 
\sqrt{s}(s \! -  \! s_0) K_1 \!    \left( \! \frac{\sqrt{s}}{T} \right) \!,~~~~
\ee
where $\sigma(s)$ is the DM annihilation cross section, $K_n$ is a modified Bessel function of the $n^\text{th}$ kind,   and
$s_0 = 4m_\chi^2$.
Starting from $n_\chi = n_\chi^{\rm eq}$ at early times, the observed relic density is achieved for $\langle \sigma v\rangle \sim 10^{-26} \rm cm^3 s^{-1}$, with the precise value depending on mass and spin, which introduce variations of order unity \cite{Steigman:2012nb}.

Since $\Ap$ couples to all charged SM species, the total annihilation cross section is a sum 
of contributions from all kinematically allowed channels, $\chi  \chi \to \bar f f$. 
For DM masses near the quantum chromodynamics confinement scale $\sim 200$ MeV, annihilation to hadronic final states
can be modeled as \cite{Izaguirre:2015yja,CarrilloGonzalez:2021lxm}
\be
\sigma_{\chi \chi \to \rm had}= R(s) \, \sigma_{\chi\chi \to \mu^+\mu^-}~,~
R(s) = \frac{ \!\!\!\! \sigma_{e^+e^-\to \rm had}}{\sigma_{e^+e^- \to \mu^+\mu^-}},~~~~
\ee
where the measured $R$-ratio is given in Ref. \cite{Ezhela:2003pp}. As we will see below, the viable mass range for freeze out via direct annihilation is approximately MeV $< m_\chi <$ GeV, so the total cross section can be written
\be
\langle \sigma v\rangle = \sum_{\ell} \langle \sigma v\rangle_{\chi \chi \to  \bar \ell \ell}
+ \langle \sigma v\rangle_{\chi \chi \to \rm had}~,
\ee
where $\ell = e,\mu,\tau$ and we only include kinematically allowed channels.
Near the resonance at $m_{\Ap} \approx 2 m_\chi$, it is necessary to include finite width effects for the $\Ap$ mediator, where 
\be
\label{eq:GammaAp}
\Gamma_{\Ap} =  \sum_f \Gamma_{\Ap \to \bar f f } + \Gamma_{\Ap \to \rm had} + \Gamma_{\Ap \to \chi \chi} ,~~
\ee
 is the total width, 
 $\Gamma_{\Ap \to \rm had } = R(m_{\Ap}) \Gamma_{\Ap \to \mu^+ \mu^-}$ is the hadronic component,
and the sum is over elementary particles with $\Gamma_{\Ap \to \bar f f} = \epsilon^2 \alpha m_{\Ap}/ 3$  in the $m_f = 0$ limit.

\begin{figure}[t!]
\hspace{-1cm}
\includegraphics[width=\linewidth]{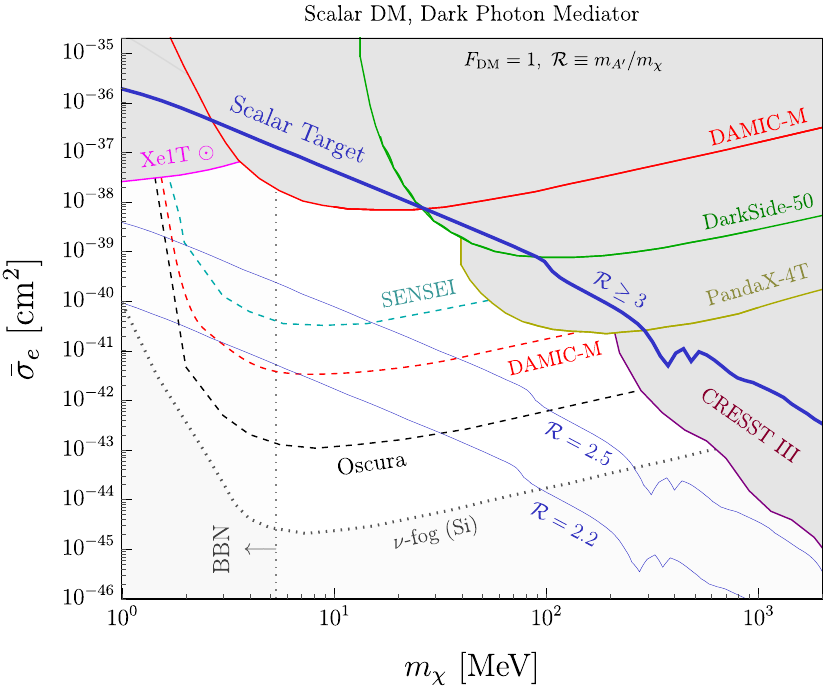} 
\caption{Limits on the effective dark matter-electron scattering cross section $\bar\sigma_e$ from Eq.~\eqref{eq:sigmae} plotted against thermal relic targets for complex scalar DM 
with different values of ${\cal R} = m_\Ap/m_\chi$. 
We also show projections for SENSEI \cite{SENSEI:2019ibb}, DAMIC-M \cite{Castello-Mor:2020jhd}, and Oscura \cite{Oscura:2022vmi}; the bottom shaded region is the 
 neutrino fog for electron recoils in Si \cite{Carew:2023qrj}.
We only include direct detection limits since these do not require any assumption about the specific values of $\alpha_D$.
The bottom region is the neutrino fog for electron scattering off a silicon target \cite{Carew:2023qrj}. Importantly, we do not show any limits from accelerator searches as these require an assumption about the specific values of $\alpha_D$ and $R$. Here we emphasize that, away from the fine-tuned resonant region at $m_{\Ap} \approx 2m_\chi$ the complex scalar model is ruled out solely based on direct detection limits. 
 }
\label{fig:main}
\end{figure}

\medskip
\noindent{\bf CMB Energy Injection.}
Although thermal DM freezes out  at  temperatures of order $T \approx m_\chi/20$ \cite{Steigman:2012nb},  annihilation reactions still occur at much lower temperatures, though they are inefficient at changing the comoving DM density.  If the annihilation cross section is $s$-wave and the comoving DM population does not change appreciably between freeze out and recombination, the resulting energy injection can modify the ionization history of the universe. Increased ionization affects CMB temperature anisotropies after recombination and can be constrained using existing data, with maximum sensitivity to energy injection around redshift $z \approx 600$ \cite{Finkbeiner:2011dx}. 
The Planck collaboration places stringent limits on the cross section \cite{Planck:2018vyg}
\be
\label{eq:cmb}
\langle \sigma v\rangle^{\rm cmb}_{\chi \chi \to \rm vis.}  \lesssim  2 \times 10^{-26} {\rm cm^3 s^{-1}}
\! \brac{0.4}{ f_{\rm eff} } \!\! \brac{m_\chi}{ 30 \rm \, GeV} \! ,~~~~~~
\ee
where vis. represents any visible SM final state, $f_{\rm eff}$ is an ionization efficiency factor that ranges approximately from $0.1-1$, depending on the final state, and asymptotes to $f_{\rm eff} \approx 0.4$ for $m_\chi \gtrsim$ GeV, assuming $\chi \chi \to e^+e^-$ annihilation \cite{Slatyer:2015jla}.

For $s$-wave annihilation, the cross-section has the same value at freeze-out as it does during 
recombination,  so models with this kind of annihilation are excluded unless the DM population changes in between freeze out and recombination.
Although the pseudo-Dirac  and asymmetric Dirac models 
undergo $s$-wave annihilation, their population changes in between chemical decoupling and recombination, so annihilation shuts off at late times and Eq.~\eqref{eq:cmb}  can
easily be satisfied.
For the complex scalar and Majorana models considered here, the annihilation is $p$-wave, so the bound in Eq.~\eqref{eq:cmb} is trivially 
satisfied since the DM velocity is significantly redshifted in between freeze out and recombination, $\langle \sigma v\rangle^{\rm cmb}_{\chi \chi \to \rm vis.} \ll 10^{-26} \rm cm^3 s^{-1}$. 

\section{Direct Annihilation Regime}
In this section we describe representative DM models that are safe
from the strong CMB limits from Eq.~\eqref{eq:cmb} and allow for the 
predictive direct-annihilation topology shown in Fig. \ref{fig:cartoon} (middle).

\begin{figure*}[t!]
\hspace{-1cm}
\includegraphics[width=0.48\linewidth]{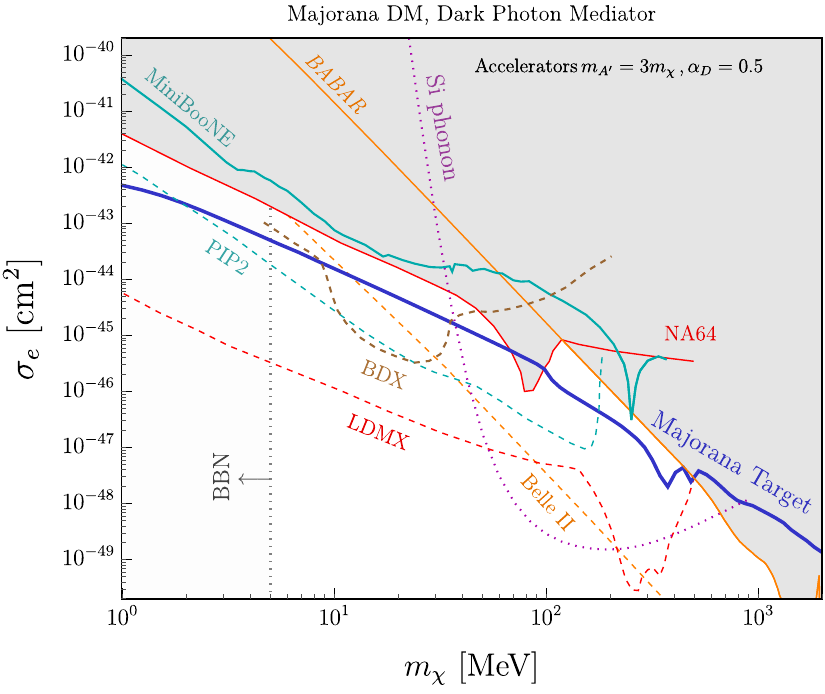}~~~
\includegraphics[width=0.48\linewidth]{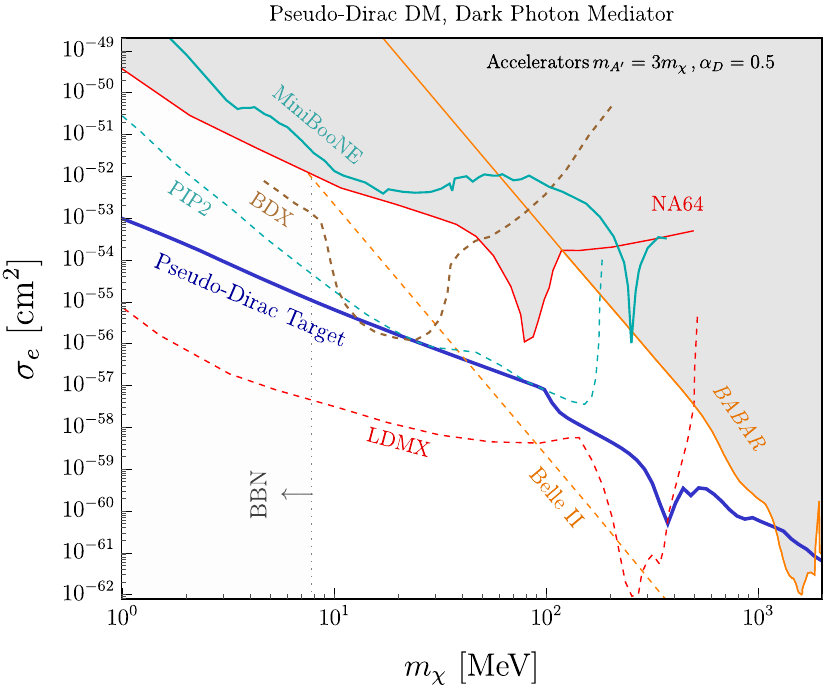}
\caption{{\bf Left:} Limits on the effective dark matter-electron scattering cross section $\bar\sigma_e$ from for Majorana DM plotted against {\it accelerator} limits from BABAR \cite{Essig:2013vha}, MiniBooNE \cite{MiniBooNEDM:2018cxm}, and NA64 \cite{NA64:2024klw}.  We also show projections for BDX \cite{BDX:2016akw}, PIP2 \cite{Dutta:2023fij}, Belle-II \cite{Izaguirre:2013uxa,Essig:2013vha}, and LDMX \cite{LDMX:2018cma}. For all accelerator curves, we adopt the conservative choices $\alpha_D = 0.5$ and $m_\Ap = 3 m_\chi$; other choices move these curves down relative to the thermal target, except near the fine-tuned resonant regime $m_\Ap \approx 2m_\chi$ \cite{Izaguirre:2015yja,Feng:2017drg}. The dotted purple curve labeled ``Si phonon"  is based on a concept for involving  phonon excitation from DM scattering off bound nuclear targets, assuming a 1 eV threshold translated into the $\sigma_e$ parameter space \cite{Kahn:2020fef}. We do not show direct detection projection for electron recoil searches because the DM form factor for this is proportional to $q^2$, which is unlike the commonly studied $F_{\rm DM} = 1$ or $1/q^2$ scenarios. Thus, the known projections for these cases cannot be reliably adapted to the Majorana case at this time. 
{\bf Right:} Same as left, but for the Pseudo-Dirac model instead.  Note that for both panels we do not show the neutrino fog shown in Fig. \ref{fig:main} as it covers much of each plane for electron recoil searches \cite{Carew:2023qrj}.
 }
\label{fig:majorana}
\end{figure*}

\medskip\noindent{\bf Complex Scalar.}
In this model,  $\chi$ couples to $\Ap$ through the current
$
J_D^\mu  = i (\chi^* \partial^\mu \chi  - \chi\partial^\mu \chi^*)$ from Eq. \eqref{eq:models},
and  the annihilation cross section for each channel is
\be
\label{eq:sv-direct-scalar}
 \sigma_{\chi \chi^* \to  \bar f f} =
  \frac{4 \pi \alpha \alpha_D \epsilon^2  \beta_\chi \beta_f   (s + 2m_\chi^2)  }{3 s [ (s - m_{A^\prime}^2)^2 +  m_{A^\prime}^2 \Gamma_{\Ap}^2 ]},
\ee
where $\beta_i =  \sqrt{ 1 - 4m_i^2/s} $.
Note that, because this process is $p$-wave, the annihilation rate is sharply suppressed at recombination, so energy injection into the CMB does not constrain the thermal-relic cross section \cite{Planck:2018vyg}. 
 The non-relativistic direct-detection cross section for scattering off target particle $f$ can be written 
\be
\label{eq:scalar-dd}
\sigma_f = \frac{16 \pi  \epsilon^2 \alpha \alpha_D \mu_{\chi f}^2}{m_{A^\prime}^4}~,
\ee
where $\mu_{\chi f}$ is the $\chi$-$f$ reduced mass and we have neglected small corrections from the momentum transfer, corresponding to the $F_{\rm DM} = 1$ form factor \cite{Essig:2011nj}.

In Fig. \ref{fig:main} we show the thermal-relic parameter space in terms of the $\chi$-$e$ scattering cross section alongside existing experimental limits and projections for future searches.  The left panel presents the $m_\Ap > 2m_\chi$ regime in which the total width satisfies  $\Gamma_\Ap \approx \Gamma_{\Ap \to \chi\chi^*}$ and 
is nearly independent of $\epsilon$. Similar results hold for the $2m_\chi > m_\Ap > m_\chi$ (not shown in Fig. \ref{fig:main}) in which the relic abundance still arises from direct annihilation (Fig. \ref{fig:cartoon}, middle), but $\Ap$ can only decay
to SM particles, so $\Gamma_{\Ap} \propto \epsilon^2$.  In both regimes, the complex scalar model is completely excluded except for the 
narrow resonance region around $m_\Ap \approx 2 m_\chi$.

\medskip\noindent{\bf Majorana.} 
In this model, the $\chi$-$\Ap$ coupling is 
$
J_D^\mu = \frac{1}{2} A^\prime_\mu \overline \chi \gamma^\mu \gamma^5 \chi ~
$
from Eq. \eqref{eq:models}.
Here the annihilation cross section has the same form as Eq.~\eqref{eq:sv-direct-scalar},
which is also $p$-wave and, therefore, safe from CMB energy injection limits \cite{Planck:2018vyg}.
The direct-detection cross section for scattering off free particle $f$ at rest is \cite{Berlin:2018bsc}
\be
\sigma_f  = \frac{8\pi \alpha \alpha_D \epsilon^2 \mu_{\chi f}^2  
(3 m_\chi^2 + 2 m_\chi m_f + m_e^2) v^2
 }{m_{A^\prime}^4 (m_\chi + m_f)^2 } ,
\ee
where $v$ is the DM velocity.

\medskip \noindent{\bf Pseudo-Dirac.} 
 In this scenario, the DM-$\Ap$ its interaction is
$J^\mu_D =  i \overline \chi_1 \gamma^\mu \chi_2  + h.c. ,$
in Eq. \eqref{eq:models}, where $\chi_1$ and $\chi_2$ are respectively the ground and excited states with corresponding masses $m_{1,2}$.  
Here the relic abundance is governed by $\chi_1 \chi_2 \to$ SM coannihilation with the cross section  
\be
\label{eq:sv-dirac}
 \sigma_{\chi_1 \chi_2 \to  \bar f f} =
  \frac{4 \pi \alpha \alpha_D \epsilon^2   (s + 2m_f^2 )   (s + 2m_\chi^2 )  }{3 s [ (s - m_{A^\prime}^2)^2 +  m_{A^\prime}^2 \Gamma_{\Ap}^2 ]} \frac{\beta_f }{\beta_\chi},~~
\ee
where we have assumed that the mass splitting $m_2 - m_1 \ll m_1 \equiv m_\chi$ so that the kinematics of freeze out are unaffected by this difference.
Although this process is $s$-wave, the excited state is generically depopulated via $\chi_2  \chi_2  \to \chi_1 \chi_1$ downscattering 
after freeze out, so coannihilation shuts off before recombination and this model is safe from CMB energy injection bounds \cite{CarrilloGonzalez:2021lxm}.

For sufficiently large mass splitting, the leading elastic contribution to the scattering cross section off point-particles $f$ arises 
from a one-loop box diagram with two $\Ap$ propagators
\cite{Berlin:2018jbm}
\be
\sigma_f = \frac{\alpha^2 \alpha_D^2  \epsilon^4  m_f^4 m_\chi^2}{  9\pi m^8_{A^\prime} }\left(11 - 60 \log\frac{m_\Ap}{m_\chi} \right)^2~, 
\ee 
which is sharply suppressed relative to all other models considered here. Correspondingly, 
as shown in Fig.~\ref{fig:majorana} (right),  the thermal target parameter space is beyond the reach of current and planned direct detection experiments,
but can still be tested with accelerator searches, as discussed below.

\medskip\noindent{\bf Asymmetric Dirac.} 
Since Dirac particles with $J_D^\mu = \bar \chi \gamma^\mu \chi$ from Eq. \eqref{eq:models} have an $s$-wave annihilation cross section, identical to the expression
in Eq. \eqref{eq:sv-dirac}, they are excluded by CMB limits for masses below $\sim$ 20 GeV \cite{Planck:2018vyg}. However, this model can remain viable 
if the DM has a sufficiently large particle-antiparticle asymmetry, so that annihilation is suppressed during the CMB era \cite{Lin:2011gj,Izaguirre:2015yja}
because the antiparticle density satisfies $n_{\bar \chi} \propto n_\chi \exp(- 0.03 \eta_\chi m_{\rm Pl} m_\chi \langle \sigma v\rangle)$, where $\eta_\chi = (n_\chi -n_{\bar \chi})/s$, $s$ 
is the entropy density, and $m_{\rm Pl}$ is the Planck mass \cite{Lin:2011gj}. Note that the dark sector is initially in chemical equilibrium so achieving the observed density with a particle asymmetry requires a
{\it larger} value of $\sigma v$ than in the symmetric case in order to annihilate away more of the symmetric population.  

\begin{figure}[t!]
\hspace{-1cm}
\includegraphics[width=0.98\linewidth]{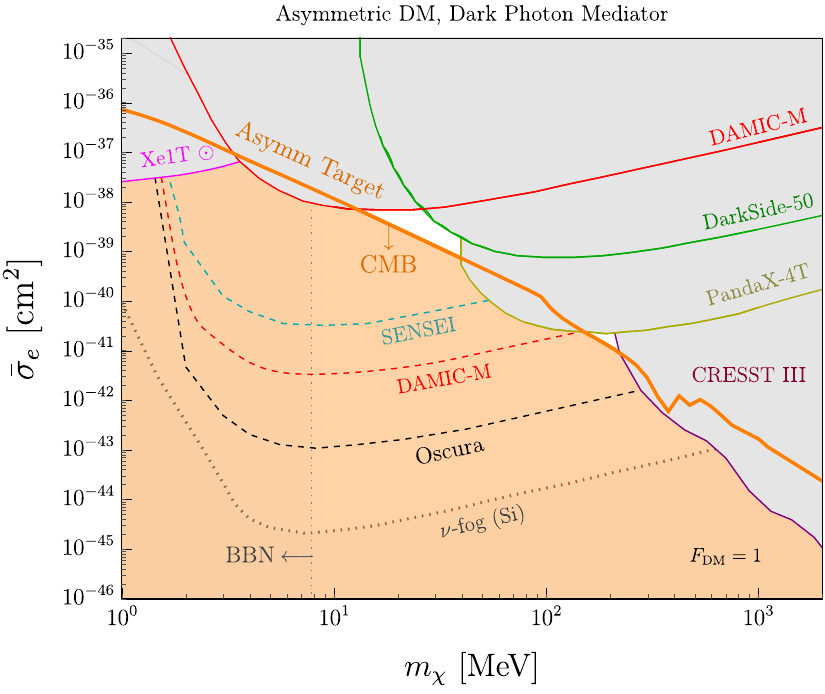}
\caption{Parameters space for asymmetric Dirac DM where gray shaded regions are excluded by
direct detection searches and the orange shaded region is excluded by CMB energy injection bounds from the residual antiparticle population at recombination. 
Unlike the other targets shown in Figs. \ref{fig:main} and \ref{fig:majorana}, here every unshaded white region is compatible with the observed DM density, though
 the DM particle-antiparticle asymmetry is different at each value point \cite{Lin:2011gj}. Here the CMB exclusion region is computed for ${\cal R} = 3$ but is insensitive
  to this ratio away from the tuned resonance region $m_\Ap \approx 2 m_\chi$. Thus, this model is now nearly excluded for parameter space off resonance.}
\label{fig:asymm}
\end{figure}

In the non-relativistic limit, the Dirac cross section for scattering off point-particle 
target $f$ is has the same form as Eq.~\eqref{eq:scalar-dd} and in Fig. \ref{fig:asymm} we 
present the viable parameter space  for this scenario. The orange shaded region  is excluded
by Planck \cite{Izaguirre:2015yja} and, unlike the other scenarios, here every point above this region
can yield the observed DM density, though each one has a different value of $\eta_\chi$. As in the complex scalar case, 
here the parameter space is nearly excluded solely by direct detection experiments.

\section{ Secluded Annihilation Regime}

Thus far, we have only considered the direct annihilation regime in which the relic density 
  depends on the SM coupling $\epsilon$. Naively, there should not be 
strong generic bounds on the ``secluded" regime ($m_\chi > m_{\Ap})$ because the $\chi\chi \to\Ap \Ap$ annihilation (Fig. \ref{fig:cartoon}, right) is independent of the SM coupling $\epsilon$ \cite{Pospelov:2007mp}. However, since $\Ap$ is the lightest particle in the dark sector, it promptly\footnote{If $\epsilon$ is sufficiently small, the $\Ap$ lifetime can be arbitrarily long. However, for $\epsilon \lesssim \sqrt{m_{\Ap}/M_{\rm Pl}}$ is the asymmetry ratio, where $M_{\rm Pl}$ is the Planck mass, the $\chi$ and $\Ap$ both lose thermal contact with the SM, so thermal freeze out is no longer viable. } decays to SM particles, so secluded annihilation is subject to the CMB bound mentioned above.

For the models considered in Eq. \eqref{eq:models}, in the secluded regime the annihilation cross sections
all satisfy  
\be
\label{eq:sigmav_secluded}
\sigma v_{\chi \chi \to \Ap \Ap} = \frac{\lambda \pi \alpha_D^2}{m_\chi^2}~~,
\ee
where we have taken the
the $m_{\chi} \gg m_{A^\prime}$ limit and  $\lambda = 1, 5/2, 11/2$ for complex scalar, Majorana fermion, and (pseudo) Dirac fermions, respectively. 
Since this expression is independent of the SM coupling $\epsilon$, the freeze out requirement  
$\sigma v_{\chi \chi \to \Ap \Ap} \sim 10^{-26} {\rm cm^3 s^{-1}}$ does not predict a testable milestone for experimental searches, whose signal rates all depend on $\epsilon$.

\section{Experimental Searches}

\noindent{\bf Electron Recoils.} 
 Electron recoil experiments probe the $\chi$-$e$ cross section, parametrized as the effective quantity \cite{Essig:2011nj}
\be
\label{eq:sigmae}
\bar \sigma_e = \frac{16\pi \alpha \alpha_D \epsilon^2 \mu_{\chi e}^2 }{  (m_{A^\prime}^2 + \alpha^2 m_e^2)^2},
\ee
which is multiplied by a material-specific structure function to calculate the total event rate. Note that a viable thermal cosmology requires $m_\Ap >$ MeV, so the models
studied here all correspond to the heavy-mediator form-factor $F_{\rm DM} = 1$ \cite{Essig:2011nj}.

The DAMIC-M collaboration has recently published new constraints on DM-$e$ interactions in a Si target with a $\sim$ 1.3 kg$\cdot$day exposure \cite{DAMIC-M:2025luv}.
The PandaX-4T experiment has placed limits on sub-GeV DM scattering off electrons  
using a xenon target with 0.55 ton$\cdot$year exposure \cite{PandaX:2022xqx}.
The DarkSide-50 collaboration has reported DM-$e$ scattering constraints in a liquid-Ar target with 12,306 kg$\cdot$day exposure, and 
these limits are all shown in Figs. \ref{fig:main} and \ref{fig:asymm}
for the complex scalar and asymmetric fermion scenarios, respectively \cite{DarkSide:2022knj}. These figures 
also show projections for
 SENSEI \cite{SENSEI:2019ibb}, DAMIC-M \cite{Castello-Mor:2020jhd}, and Oscura \cite{Oscura:2022vmi}. 
Electron recoil sensitivity can also be enhanced at low mass through the process
of solar reflection, in which a small fraction of DM particles are upsacttered to higher velocities through their interactions 
with the sun, and can thereby deposit more energy in terrestrial detectors \cite{An:2021qdl}.

\medskip
\noindent{\bf Nuclear Recoils.} For an $\Ap$ mediator, 
nuclear recoil searches constrain the cross section
for scattering off detector protons
\be
\label{eq:sigmap}
\sigma_p = \frac{16\pi \epsilon^2 \alpha_D \alpha \mu^2_{\chi p}}{m_{A^\prime}^2}  = \frac{\mu^2_{\chi p}}{ \mu^2_{\chi e}} \bar \sigma_{\chi e}^2,
\ee
where $\bar \sigma_e$ is defined in Eq. \eqref{eq:sigmae}.
In Figs. \ref{fig:main} and \ref{fig:asymm} we show CRESST-III limits from Ref. \cite{CRESST:2019jnq,CRESST:2024cpr}
which are based on a 3.64 kg$\cdot$day exposure with a CaWO$_4$ target; here we 
translate $\sigma_p$ into the $\bar \sigma_e$ using Eq.~\eqref{eq:sigmap} and note that the limits in Ref. \cite{CRESST:2019jnq,CRESST:2024cpr} assume that DM couples to all nucleons in the target, but for an $\Ap$ mediator it only couples to the protons. 

Nuclear recoil search sensitivity can be extended down to lower mass through the Migdal effect \cite{Migdal1939} in which DM-nucleus
interactions catalyze atomic ionization \cite{Ibe:2017yqa,Dolan:2017xbu,Bell:2019egg,Kahn:2020fef,Baxter:2019pnz,Knapen:2020aky}. 
Due to the difficulty of calibrating this phenomenon \cite{PhysRevD.109.L051101}, in Fig. \ref{fig:main} we conservatively omit limits based the Migdal effect, even though such searches have been reported by several experiments \cite{DarkSide:2022dhx,PandaX:2023xgl,SENSEI:2023zdf,SuperCDMS:2023sql,XENON:2019zpr}. However, this situation could change significantly as there are several proposals to calibrate this effect for dark
matter detection in the near future \cite{MIGDAL:2022yip,Adams:2022zvg}. 
Related ideas for detecting phonon excitation
energy from nuclear targets may also be promising \cite{Schutz:2016tid,Trickle:2020oki,Kahn:2020fef}.

\medskip
\noindent{\bf Big Bang Nucleosynthesis.}
For $\lesssim 10$ MeV scale thermal DM, freeze out occurs around neutrino decoupling, 
just before big bang nucleosynthesis (BBN). The resulting modification to the Hubble rate and 
the entropy transfer to SM species affect the light element yields formed during this epoch. In Figs. \ref{fig:main}-\ref{fig:asymm} we show the 
bound on light thermal DM based on the successful predictions of standard BBN \cite{Nollett:2013pwa,Boehm:2013jpa,Krnjaic:2019dzc,Sabti:2019mhn}

\medskip
\noindent{\bf Beam Dumps.} Light DM can be produced relativistically at accelerator beam dumps and scatter of SM particles to deposit energy in a downstream detector \cite{deNiverville:2011it}. Currently the the MiniBooNE experiment places the best limits on sub-GeV DM using proton beams impinging on a fixed target to produce the DM flux \cite{Aguilar-Arevalo:2019wki,MiniBooNEDM:2018cxm,MiniBooNE:2017nqe}. Future searches using the proposed DUNE-PRISM \cite{DeRomeri:2019kic}, SHiP \cite{SHiP:2020noy}, and PIP2 \cite{Dutta:2023fij} experiments are poised to significantly improve coverage of this parameter space.

Electron beam dump searches are also a promising strategy for testing light DM \cite{Izaguirre:2013uxa}. Currently, the best limit of this kind is from the 
E137 experiment \cite{Batell:2014mga}, but, as shown in Fig. \ref{fig:majorana}, this has since been superseded by other kinds of constraints. However, the planned BDX experiment
can greatly improve upon E137 and cover important thermal milestones in this parameter space \cite{BDX:2016akw,Battaglieri:2022dcy}

\medskip
\noindent{\bf B-Factories.}
Low energy $e^+e^-$ colliders are a well known probe of dark photons coupled to light DM. 
Existing BABAR searches have been interpreted
to place limits on $\epsilon$ through the $e^+ e^- \to \gamma \Ap$ where the $\Ap \to \chi \chi$  decay yields missing energy \cite{Izaguirre:2013uxa,Essig:2013vha}.
Future searches with Belle-II are poised to improve these limits with the full 50 ab$^{-1}$ data set, which the experiment is expected to collect over its full runtime. 

\medskip
\noindent{\bf Missing Energy/Momentum.}
Electron-beam missing energy experiments can probe invisibly decaying 
$\Ap$ through the radiative process $e N \to e N \Ap$, where $N$ is a target nucleus  embedded in an active, instrumented target. For this strategy, the kinematics of individual beam particles are measured both upstream and downstream of the interaction to detect anomalous energy loss, interpreted as $\Ap$ production followed by a decay to DM particles. As shown in Fig.~\ref{fig:majorana}, the NA64 experiment
has placed the strongest limits on light DM using this technique \cite{NA64:2023wbi}. 

The related missing-momentum strategy studies the same radiative production process, but uses a thin target, such that tracking layers can monitor the transverse momentum of the recoiling electron beam in order to infer the production of invisible DM particles \cite{Izaguirre:2014bca}. In Fig.~\ref{fig:majorana} we show projections for phase-II of the 
the proposed LDMX experiment, which is based on the missing momentum strategy \cite{LDMX:2018cma,Akesson:2022vza}.

\medskip
\noindent{\bf Indirect Detection.}
Since CMB energy-injection limits place strong bounds on any model with $s$-wave annihilation, all the  models here feature sharply suppressed annihilation rates around the epoch of recombination. For complex scalar and Majorana DM, the annihilation
cross sections are $p$-wave, $\sigma v \propto v^2 $,  so  $\sigma v^{\rm cmb}_{\chi\chi \to \bar f f} \ll \sigma v^{\rm freeze-out}_{\chi\chi \to \bar f f}$ and 
the late time cross section is beyond the reach of existing or proposed search strategies. 

Although the  pseudo-Dirac scenario predicts an $s$-wave coannihilation cross section, this model can still be safe from CMB limits for a sufficiently large mass splitting between the two DM states \cite{CarrilloGonzalez:2021lxm}. In this model, freeze out proceeds via $\chi_1 \chi_2 \to \bar f f$ coannihilation, but
the downscattering process 
$\chi_2 \chi_2\to \chi_1 \chi_1$ can remain highly efficient after $\chi_{1,2}$ chemically decouple from the SM. Thus, during the CMB era, the $\chi_2$ number density 
is exponentially suppressed and $\chi_1 \chi_2$ coannihilation can be sufficiently rare to evade Planck limits \cite{Planck:2018vyg}. However, at late times when structure forms, $\chi_1$ particles trapped in gravitational potentials
can convert their kinetic energy to locally revive the $\chi_2$ population via $\chi_1 \chi_1 \to \chi_2 \chi_2$ reactions. This $\chi_2$ population can then coannihilate with the ambient $\chi_1$ particles via $\chi_1 \chi_2 \to \bar f f$ reactions, which yield a secondary photon flux  \cite{Berlin:2023qco}.

\section{\bf Discussion} 
The experimental study of sub-GeV DM is now sufficiently mature to test key milestones
for thermal DM production in the early universe. In particular, several experiments are now probing the direct annihilation regime, which predicts a one-to-one correspondence between the DM-SM scattering cross section and the direct-annihilation cross section, which governs the relic density. Here we have surveyed a broad range of minimal thermal models with a dark photon mediator and found that the complex scalar model is officially ruled out solely based on recent direct-detection constraints. Future direct detection experiments will play a key role in probing the resonant annihilation regime and approaching sensitivity to the Majroana relic target. However, dedicated $B$-factory and fixed-target accelerator searches are necessary to fully probe all of the (non-resonant) thermal targets studied here.

Finally, although our analysis is specific to a kinetically-mixed dark photon mediator, there is nothing special about this choice. Every scenario studied here extends almost identically
to corresponding scenarios whose vector mediators are the gauge bosons of anomaly-free abelian extensions to the SM \cite{Ilten:2018crw,Bauer:2018onh}. As long as the mediator couples to the first generation, nearly all of the thermal targets, experimental bounds, and future projections also apply to these scenarios with only order-one variations; the only exception is 
gauged $\mu - \tau$ flavor, for which the electron and proton couplings are significantly suppressed relative to those of the dark photon \cite{Kahn:2018cqs}.

{\bf Acknowledgments.}
We thank Asher Berlin, Nikita Blinov, Yoni Kahn, and Tanner Trickle  for helpful conversations. 
Special thanks to Dan ``DBax" Baxter for helpful conversations and encouraging this paper to be written.
This manuscript has been authored in part by  Fermi Forward Discovery Group, LLC under Contract No. 89243024CSC000002 with the U.S. Department of Energy, Office of Science, Office of High Energy Physics. This research was supported in part by grant NSF PHY-2309135 to the Kavli Institute for Theoretical Physics (KITP).

\bibliography{biblio}
\end{document}